\documentclass[twocolumn,showpacs,preprintnumbers,amsmath,amssymb,
prl]{revtex4}

\usepackage{graphicx}
\usepackage{dcolumn}
\usepackage{bm}
\usepackage{epsf,epsfig,latexsym}

\begin{document}
\title{Experimental determination of the symmetry energy of a low
density
nuclear gas}

\author{S. Kowalski}
\thanks{Now at Institute of Physics, Silesia University, Katowice,
Poland}
\author{J. B. Natowitz}
\author{S. Shlomo}
\author{R. Wada}
\author{K. Hagel}
\author{J. Wang}
\thanks{Now at Institute of Modern Physics Chinese Academy of
Science, Lanzhou 73, China}
\author{T. Materna}
\author{Z. Chen}
\author{Y. G. Ma}
\thanks{Now at  Shanghai Institute of Nuclear Research, Chinese
Academy of
Sciences, Shanghai 201800, China}
\author{L. Qin}
\author{A. S.  Botvina}
\thanks{Now at Institute for Nuclear Research, Russian Academy of
Sciences,
117312 Moscow, Russia}
\affiliation{Cyclotron Institute, Texas A\&M University, College
Station,
Texas 77843}
\author{D. Fabris}
\author{M. Lunardon}
\author{S. Moretto}
\author{G. Nebbia}
\author{S. Pesente}
\author{V. Rizzi}
\author{G. Viesti}
\affiliation{INFN and Dipartimento di Fisica dell' Universit\'a di
Padova,
I-35131 Padova, Italy}
\author{M. Cinausero}
\author{G. Prete}
\affiliation{INFN, Laboratori Nazionali di Legnaro, I-35020
Legnaro, Italy}
\author{T. Keutgen}
\author{Y. El Masri}
\affiliation{FNRS and IPN, Universit\'e Catholique de Louvain,
B-1348 Louvain-la-Neuve, Belgium}
\author{Z. Majka}
\affiliation{Jagellonian University, M Smoluchowski Institute of
Physics,
PL-30059, Krakow, Poland}
\author{A. Ono}
\affiliation{Department of Physics, Tohoku University, Sendai,
Japan}

\date{\today}

\begin{abstract}
Experimental analyses of moderate temperature nuclear gases produced in the
violent collisions of 35 MeV/nucleon$^{64}$Zn projectiles with $^{92}$Mo and
$^{197}$Au target nuclei reveal a large degree of alpha particle clustering
at low densities.  For these gases, temperature and density dependent
symmetry energy coefficients have been derived from isoscaling
analyses of the yields of nuclei with A $\leq 4$.  At densities of 0.01 to
0.05 times the ground state density of symmetric nuclear matter, the
temperature and density dependent symmetry energies range from
9.03 to 13.6 MeV.  This is much larger than those obtained in mean field
Calculations and reflects the clusterization of low density nuclear
matter.  The results are  in quite reasonable agreement with
calculated values obtained with a recently proposed Virial
Equation of State calculation.
\end{abstract}

\maketitle

In a recent theoretical paper, Horowitz and Schwenk have reported
the development of  a Virial Equation of State (VEOS) for low density
nuclear matter~\cite{horowitz05}. This virial equation of state
is a thermodynamically consistent equation of state, in which the
virial coefficients for nucleon-nucleon, nucleon-alpha and alpha-
alpha interactions are derived directly from experimental binding
energies and scattering phase shifts. It is argued that
contributions from $^3$H and $^3$He are expected to be small and they are
ignored in the calculation. In the work reported in reference
~\cite{horowitz05} these virial coefficients were then used to
make predictions for a variety of properties of nuclear matter
over a range of density, temperature and composition.  The authors
view this virial equation of state, derived from experimental
observables, as "model-independent, and therefore a benchmark for
all nuclear equations of state at low densities.  Its importance
in both nuclear physics and in the physics of the neutrino sphere
in supernovae is discussed in the VEOS paper~\cite{horowitz05}.
A particularly important feature of the VEOS, emphasized in
reference~\cite{horowitz05}, is the natural inclusion of
clustering which leads to large symmetry energies at low baryon
density.

In this paper we extend our investigations of the nucleon and
light cluster emission that occurs in near-Fermi energy heavy ion
collisions~\cite{hagel00,wada04,wang05_1,wang05_2,wang06} to
investigate the properties of the low density participant matter produced 
in such collisions.  The data provide experimental evidence for a large 
degree of alpha clustering in this low density matter, in agreement with 
theoretical predictions~\cite{beyer00,beyer04,shen98,horowitz05}.  Temperature
and density dependent symmetry free energies and symmetry energies have been
determined at densities of $0.05\rho_0$ or less, where $\rho_{0}$ is the
ground state density of symmetric nuclear matter, by application of an
isoscaling analysis~\cite{tsang01_1,tsang01_2}.  The symmetry energy
coefficient values obtained, 9.03 to 13.6  MeV, are much larger then
those derived from effective interactions in mean field models.
The values are in reasonable agreement with those calculated in the
VEOS treatment of reference~\cite{horowitz05}.

\section{Experimental Procedures}
The reactions of 35A MeV $^{64}$Zn projectiles with $^{92}$Mo and $^{197}$Au
target nuclei were studied at the K-500 Super-Conducting Cyclotron
at Texas A\&M University, using the 4$\pi$ detector array
NIMROD~\cite{wada04}. NIMROD consists of a 166 segment charged
particle array set inside a neutron ball. The charged particle
array is arranged in 12 concentric rings around the beam axis. In
those rings, the individual segments are fronted by ionization
chambers (IC) filled with 30 Torr of CF$_4$ gas. Front and back
windows were made of 2.0 $\mu$m aluminized Mylar foil. In each of
these forward rings, two of the segments have two Si detectors(
150 and 500 $\mu$m thick) between the IC and CsI detectors (super
telescopes) and three have one Si detector(300 $\mu$m thick). Each
super telescope is further divided into two sections. The CsI
detectors are 10 cm thick Tl doped crystals read by
photomultiplier tubes. For these detectors, a pulse shape
discrimination method is employed to identify light particles. In
all telescopes particles are identified in atomic number. In the
super telescopes, all isotopes with atomic number Z=10 are clearly
identified.

The energy calibration of the Si detectors was carried out using
alpha particles from  a $^{228}$Th source and the observed punch
through energies of identified reaction products. Since the energy
losses of the lighter particles, in particular the high energy
Hydrogen isotopes, are rather small in the Si detectors,
evaluation of the energy deposited in the CsI crystal from the
energy loss in the Si detectors requires special care for higher
energy particles. Therefore, an additional energy calibration was
performed employing Si detectors of thicknesses 1mm, backed by CsI
detectors of three different lengths (1cm, 3cm and 5cm) to measure
the inclusive energy spectra of light charged particles from the
reaction $^{64}$Zn + $^{92}$Mo at 47A MeV  The energy spectra were measured
at all angles corresponding to those of the 12 rings of NIMROD.
The combination of thicker Si E detectors and observation of high
energy punch-through points for the particles which traversed
these thinner CsI detectors allowed us to determine the energy
spectra with a high degree of confidence. We then used the $^{64}$Zn +
$^{92}$Mo at 47A MeV as a standard reaction to determine the CsI energy
calibrations for all other runs. Neutron multiplicity was measured with
the 4$\pi$ neutron detector surrounding the charged particle array.
This detector, a neutron calorimeter filled with Gadolinium doped
pseudocumene, consists of two hemispherical end caps and a
cylindrical mid-section. The mid-section is divided into four
separate 90 degree quadrants. The hemispheres are 150 cm in
diameter with beam pipe holes in the center and they are upstream
and downstream of the charged particle array. Thermalization and
capture of emitted neutrons in the ball leads to scintillation
which is observed with phototubes providing event by event
determinations of neutron multiplicity but little information on
neutron energies and angular distributions. Further details on the
detection system, energy calibrations and neutron ball efficiency
may be found in reference~\cite{wada04}.

\section{Data Analysis}
An inspection of the two dimensional arrays depicting the detected
correlation between charged particle multiplicity and neutron
multiplicity in NIMROD (not shown), reveals a distinct correlation
in which increasing charged particle multiplicity is associated
with increasing neutron multiplicity. Although there are
significant fluctuations reflecting both the competition between
different decay modes and the neutron detection efficiencies,
these correlations provide a means for selecting the higher
multiplicity, more violent collisions. For the analysis reported
in this paper, we have selected events corresponding to the
largest observed neutron and charged particle multiplicities. For
$^{64}$Zn+$^{197}$Au at 35 AMeV this selection corresponded to 442,000 events,
while for  $^{64}$Zn+$^{92}$Mo at 35 AMeV            it corresponded to 280,000
events.  Many of the techniques applied in this analysis have been
discussed previously in greater detail in references
~\cite{wada04,wang05_1,wang05_2,wang06}.  The detected
correlation between charged particle multiplicity and neutron multiplicity in
NIMROD provides a means for selecting the collision
violence~\cite{wada04,wang05_1,wang05_2,wang06}.    For the systems studied,
reference~\cite{wada04} also contains an extensive comparison of
earlier experimental results with predictions of Antisymmetrized Molecular
Dynamics calculations (AMD-V)~\cite{ono99}

For the selected events we carried out analyses using three-source fits to
the observed energy and angular distributions of the light charged particles.
The assumed sources are the projectile-like fragment source (PLF), the
target-like fragment source (TLF) and an intermediate velocity (IV)
source~\cite{awes81_1,awes81_2,prindle98,wada89}.  Even though the system
evolves in a continuous fashion, such source fits provide a useful schematic
picture of the emission process.  From the fits we obtained parameters
describing the ejectile spectra and multiplicities that can be associated to
the three different sources.  As in the earlier works, the IV source is
found to have a source velocity very close to half of that of the projectile
reflecting the initial decoupling of the momentum sphere of the participant
matter from that of the remaining nucleons.  This important feature of the
dynamically evolving system manifests itself as kinematic differences between
the early emitted light ejectiles (gas) and the remaining  matter (liquid).
In the following we probe the properties of this early ejectile gas as the
system relaxes toward equilibrium and the two momentum spheres become more and
more similar.

Treatment of the early cluster emission in a coalescence 
framework~\cite{csernai86,mekjian78} has proven to be a very effective tool
for understanding the energy spectra of the light 
ejectiles~\cite{awes81_1,awes81_2}.
We have previously developed extensions of these techniques to probe the
early interaction zone and dynamic evolution in Fermi-energy collisions
~\cite{cibor01,hagel00,cibor00,wang06}.  For the present work we have derived
information on the early thermalization stage of the reaction
~\cite{cibor01,hagel00,cibor00,wang05_1} by focusing on the properties of
early emitted mid-rapidity particles associated with the IV source.  Such a
selection minimizes contributions from the other sources.  In addition,
yields assigned to the TLF source are subtracted from the experimental
yields.  Thus, the yields of higher energy particles are relatively
uncontaminated by later emission processes.  AMD-V calculations reported
previously~\cite{wada04,wang05_1} indicate that the velocities of early
emitted light particles decrease rapidly with increasing average emission
time.  We have examined the evolution of the system by determining various
parameters characterizing the ejectile yields, i.e., temperature, alpha mass
fraction, N/Z ratios and isoscaling parameters, as a function of ejectile
velocity.  The velocity employed is the ``surface velocity'', V$_{surf}$, of
the emitted particles in the IV frame, defined as the velocity of an emitted
species at the nuclear surface, prior to acceleration in the Coulomb
field~\cite{awes81_1}.  The energy prior to Coulomb acceleration is
obtained in our analysis by subtraction of the Coulomb barrier energy
derived from the source fits.  In earlier studies we have employed the
calculated correlation from AMD-V predictions to calibrate the time-scale
associated with our data~\cite{cibor00,hagel00,wang05_1,wang05_2}.

\section{Temperature Determinations}
To characterize the temperature evolution of the system we have
used the
yields of $^{2}$H, $^{3}$H,$^{ 3}$He and $^{4}$He clusters to
determine the
double isotope temperature, T$_{HHe}$, as a function of V$_{surf}$
in the
IV frame~\cite{albergo85,kolomiets97}.

\begin{equation}
T_{HHe} = \frac{14.3}{\ln{(\sqrt{(9/8)}(1.59 R_{Vsurf}))}}
\label{eqnT}
\end{equation}

where R$_{Vsurf.}$, the double ratio of cluster yields, Y, for
clusters with the same surface velocity, is [Y($^2$H) Y($^4$He)] / [Y($^3$H)
Y($^3$He)] and the constants 14.3 and 1.59 reflect binding energy, spin, masses
and mass differences of the ejectiles.  Equation~\ref{eqnT}, for particles
with the same V$_{surf}$, differs from the usual formulation by a
factor of $\frac{9}{8}^\frac{1}{2}$ appearing in the logarithmic term in the
denominator~\cite{wang05_1}.

\begin{figure}
\epsfig{file=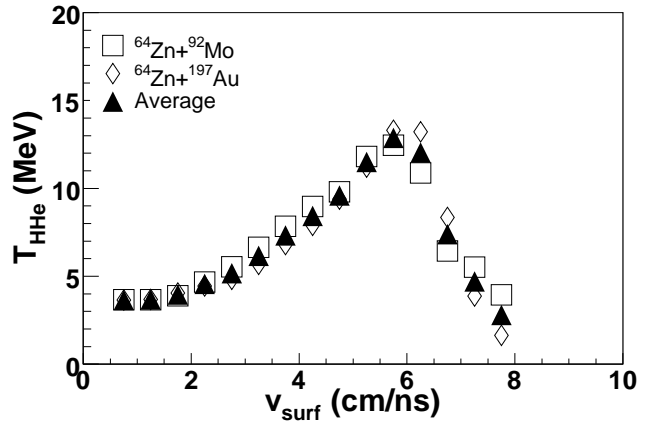,width=9.2cm,angle=0}
\caption{Double isotope ratio temperatures, T$_{HHe}$, as a function of 
surface velocity for 35 MeV/nucleon $^{64}$Zn + $^{92}$Mo (open squares)
and $^{64}$Zn +$^{197}$Au (open diamonds).  Also shown is the average of the
two (solid triangles). Uncertainties are estimated to be $\leq 30\%$}
\label{fig1}
\end{figure}

We present, in Figure~\ref{fig1}, results for the double isotope
ratio temperatures as a function of surface velocity in the intermediate
velocity frame.  These temperatures have previously been reported in
reference~\onlinecite{wang05_1}.  The temperature evolution of the
IV source ejectiles is seen to be quite similar for the two different
systems.  For each system investigated the double isotope ratio temperature
determination exhibits a high maximum temperature, near 13 MeV.  In each case,
after reaching a maximum the temperature then decreases monotonically as
the velocity decreases.  In references~\onlinecite{wang05_1,wang05_2}
we discuss this evolution and present varying pieces of evidence that at
velocities smaller than the peak velocity, the system attains chemical 
equilibrium, at least on a local basis. This appears to occur at times near 
115 fm/c.  Entry into the evaporation stage of the reaction occurs at 
V$_{surf}$ near 3 cm/ns, corresponding to times near 160 fm/c.  At that 
latter point, the temperatures are near 6 MeV, comparable to those observed 
in the caloric curve plateaus for these systems~\cite{natowitz02_1}.  In the 
plateau region average densities (liquid plus gas) as low as  $0.4\rho_0$ 
are indicated by a previous expanding Fermi-gas analysis~\cite{natowitz02_2}.

\section{Alpha Mass Fractions}
The VEOS paper~\cite{horowitz05} employs the alpha mass fraction, X$_{A}$,
defined as the ratio of mass bound in alpha particles to the total mass of
the matter under consideration, to characterize the degree of alpha
clustering at different densities and temperatures.  We have determined
experimental values of X$_{A}$ as a function of velocity.  For this purpose
it was assumed that the unmeasured neutron multiplicity at a given velocity
was the product of the $^{3}$H/$^{3}$He yield ratio times the proton yield
for that velocity.  For the IV souce particles this is implicit in coalescence
assumptions~\cite{csernai86,mekjian78,cibor01} and consistent with
experimental results~\cite{keutgen,famiano06}.  In this way both the total
mass yields and the mass fractions could be deduced.  The sum of the mass
emitted from the IV source is $31.9\pm2.5$ amu for the Mo target and
$29.7\pm2.4$ amu for the Au target.  As in the case of the temperature
evolution, the close correspondence of ejected mass provides evidence of
production of a similar zone of participant matter in the two different cases.
\begin{figure}
\epsfig{file=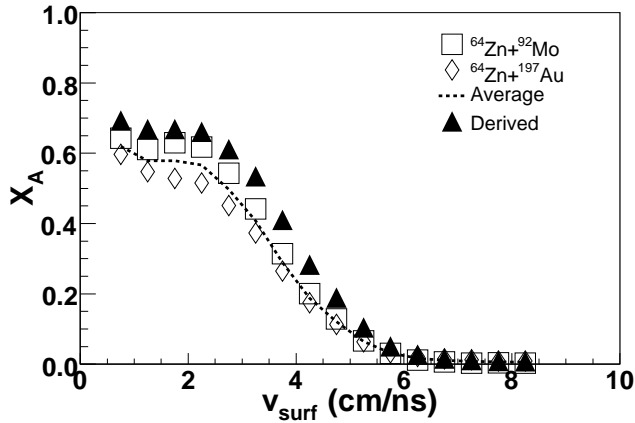,width=9.2cm,angle=0}
\caption{Alpha mass fractions, X$_{A}$, as a function of surface
velocity for 35 MeV/nucleon $^{64}$Zn + $^{92}$Mo (squares)  and
$^{64}$Zn +$^{197}$Au (diamonds). Also shown by the dashed line is the
average of the two. Solid triangles represent derived values used for
comparison to VEOS results.  See text.}
\label{fig2}
\end{figure}

In Figure~\ref{fig2} we present the results for alpha mass fractions for the
IV source ejectiles of both systems.  For the two systems X$_{A}$ evolves in
a similar fashion with surface velocity.  As the surface velocity decreases
X$_{A}$ increases dramatically.  X$_{A}$ is smaller for the more neutron rich
$^{64}$Zn +$^{197}$Au entrance channel.  Figures~\ref{fig1} and~\ref{fig2}
clearly indicate that large alpha mass fractions are associated with the
lower temperatures.

\section{Isoscaling}
Horowitz and Schwenk have pointed out that extensive alpha clustering in the
low density gas leads naturally to an increase in the symmetry energy for the 
clustering system~\cite{horowitz05}.  For comparison to the symmetry energy 
predictions of the VEOS model we have derived symmetry energies from the 
ejectile yield data by employing an isoscaling analysis.  Such analyses have 
been reported in a number of recent 
papers~\cite{tsang01_1,tsang01_2,botvina02,ono03,souliotis04,lefevre05}.  In
this approach the yields of a particular species Y(N,Z) from two
different equilibrated nuclear systems, 1 and 2, of similar temperature but
different neutron to proton ratios, N/Z,  are expected to be related
through the isoscaling relationship

\begin{eqnarray}
\frac{Y_2}{Y_1}& =& Ce^{((\mu_2(n) - \mu_1(n))N + (\mu_2(p) -
\mu_1(p))Z)/T} \\
&=& Ce^{\alpha N+\beta Z}
\end{eqnarray}

where C is a constant and $\mu(n)$ and $\mu(p)$ are the neutron and proton
chemical potentials.

The isoscaling parameters  $\alpha = (\mu_{2}(n) - \mu_{1}(n))/T$
and $\beta = (\mu_{2}(p) -\mu_{1}(p))/T$, representing the difference
in chemical potential between the two systems, may be extracted from suitable
plots of yield ratios.  Either parameter may then be related to the
symmetry free energy, F$_{sym}$.  With the usual convention that system 2 is
richer in neutrons than system 1,

\begin{eqnarray}
\alpha &=& 4F_{sym}((Z_1/A_1)^2 - (Z_2/A_2)^2) /T
\label{eqnAlpha1} \\
\beta  &=& 4F_{sym}((N_1/A_1)^2 - (N_2/A_2)^2) /T
\label{eqnBeta1}
\end{eqnarray}

where Z is the atomic number and A is the mass number of the emitter.  Thus, 
F$_{sym}$ may be derived directly from determinations of system temperatures,
Z/A ratios, and isoscaling parameters.  We emphasize that the present analysis 
is carried out for light species characteristic of the nuclear gas rather
than, as in most previous analyses, for the intermediate mass fragments
thought to be characteristic of the nuclear liquid.

In previous works attempts were made to employ systems of similar Z and 
derive the $\alpha$ parameter which is expected to be less sensitive to 
Coulomb effects.  It is normally assumed that all contributions to
F$_{sym}$ other than that arising from the symmetry energy, e.g. bulk,
Coulomb and surface, may be
ignored~\cite{tsang01_1,tsang01_2,botvina02,ono03,souliotis04,lefevre05}.
Most commonly, the expression,
\begin{equation}
\alpha = 4\gamma((Z_1/A_1)^2 - (Z_2/A_2)^2)/T
\label{eqnAlpha2}
\end{equation}
in which $\gamma$ is the symmetry energy coefficient, has been
employed with
estimates of  Z$_1$/A$_1$, Z$_2$/A$_2$ and T to obtain estimates
of
$\gamma$~\cite{tsang01_1,tsang01_2,botvina02,ono03,souliotis04,lefevre05}.
While the approximate equation~\ref{eqnAlpha2} may be reasonable
for systems near normal density, it becomes increasingly less accurate at low 
densities where entropic effects become increasingly more 
important~\cite{davila05}.  Indeed a recent paper has demonstrated the 
existence of isoscaling effects in a non-interacting system with no symmetry 
energy, resulting entirely from maximal entropy~\cite{davila05}.  Further, 
the VEOS treatment indicates that clustering in low density matter leads to 
entropic contributions to the free energy which can differ significantly in 
magnitude and, even in sign, from those obtained from mean field 
calculations~\cite{horowitz05}.  In this work we employ 
equation~\ref{eqnAlpha1} with experimentally determined isoscaling parameters,
temperatures and Z/A ratios to determine the symmetry free energy, F$_{sym}$. 
We believe the present work to be the first isoscaling analysis in which all 
relevant parameters of equations~\ref{eqnAlpha1} and~\ref{eqnBeta1} have been 
determined experimentally.  We then employ calculated entropic corrections
from reference~\cite{horowitz05} to obtain the symmetry energy, E$_{sym}$

\section{Isoscaling Parameters, $\alpha$ and $\beta$}
\begin{figure}
\epsfig{file=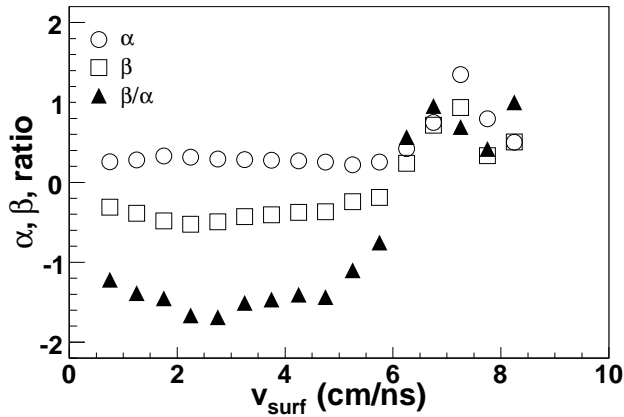,width=9.2cm,angle=0}
\caption{
Isoscaling parameters as a function of surface velocity.
The values for the parameters $\alpha$ (open circles), $\beta$ (open
squares and the parameter ratio $\alpha/\beta$ (solid triangles).  Estimated 
errors are $\pm5\%$}
\label{fig3}
\end{figure}

The isoscaling analysis of  the two systems was done in the usual
way by plotting isotope or isotone yield ratios, in our case  for Z=1 and
Z=2 particles, and doing a global fit to the yield ratios to obtain
$\alpha$ and $\beta$.  In Figure~\ref{fig3}, the isoscaling parameters,
$\alpha$ and $\beta$, are plotted against surface velocity.

We see that below 5.0 cm/ns the values vary slowly.  Below 5 cm/ns
the ratio, $\alpha/\beta$, also varies slowly but increases dramatically at
higher velocities, even changing sign.  We have previously suggested that
equilibration is observed for velocities below those for which the
peak temperatures are seen.  In our opinion the behavior of the
individual parameters and of the $\alpha/\beta$ ratio provides further
evidence that equilibration is achieved for surface velocities below 5 cm/ns.

\section{Z/A Ratios}
\begin{figure}
\epsfig{file=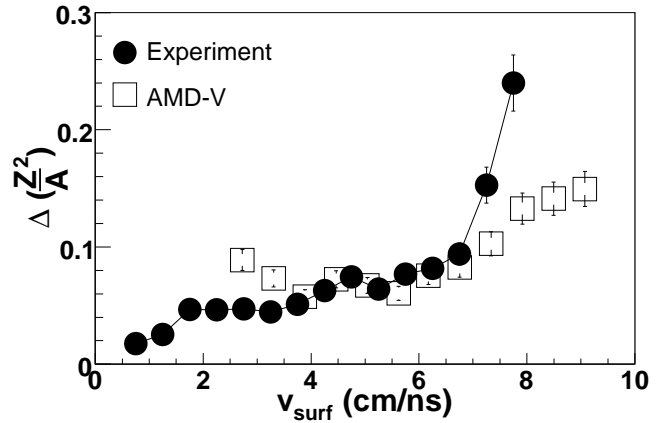,width=9.2cm,angle=0}
\caption{The quantity ((Z$_1$/A$_1$)$^2$-
(Z$_2$/A$_2$)$^2$)$^2$
as a function of surface velocity.  Experimental determinations
are
represented by solid circles.  Results of AMD-V calculations are
represented
by open squares.}
\label{fig4}
\end{figure}

For each emitting system, the sums of Z and A for the emitted
ejectiles, taken to be characteristic of that of the sampled gas, have been
derived as a function of surface velocity.  From these the values of
((Z$_1$/A$_1$)$^2$- (Z$_2$/A$_2$)$^2$)$^2$ were then obtained.
These values are presented in Figure~\ref{fig4} as a function of surface
velocity.  For comparison, theoretical values obtained for the gas fraction
derived from the AMD-V calculation are also presented.  The experimental
results and those for the gas in the dynamical AMD-V calculation are 
generally quite close in the 3 to 7 cm/ns surface velocity.

\section{Symmetry Free Energies and Densities}
Experimentally determined values of T, Z/A and $\alpha$ were used
in equation~\ref{eqnAlpha1} to calculate F$_{sym}$ as a function of
surface velocity. (Note that both temperature and density are varying with
surface velocity.) The values of F$_{sym}$ are presented in column 3 of
Table~\ref{tab1}.  Values of range from 12.2 MeV to 5.4 MeV as the
surface velocity decreases from 4.75~cm/ns to 0.75~cm/ns.

While densities are not easily accessible experimental quantities
in collision studies, knowledge of the densities at which the F$_{sym}$
determinations are being made is critical to an interpretation of
the measured values.

As pointed out by Albergo {\it et al,} knowledge of the temperature
allows the extraction of the free proton densities from the yield
ratios of ejectiles which differ by one proton, e.g., the yield
ratio of  $^4$He to $^3$He. Specifically ,

\begin{equation}
\rho_p = 0.62\times10^{36}T^{3/2}e^{-19.8/T}{Y(^4{\rm He})/Y(^3{\rm H})}
\end{equation}

Here T is the temperature in MeV,  Y refers to the yield of the
species under consideration and $\rho_p$ has units of nucleons/$cm^3$.

Correspondingly, the free neutron densities may be extracted from
the yield ratios of ejectiles which differ by one neutron.  Specifically,

\begin{equation}
\rho_n = 0.62\times10^{36}T^{3/2}e^{-20.6/T}{Y(^4{\rm He})/Y(^3{\rm He})}
\end{equation}

\begin{figure}
\epsfig{file=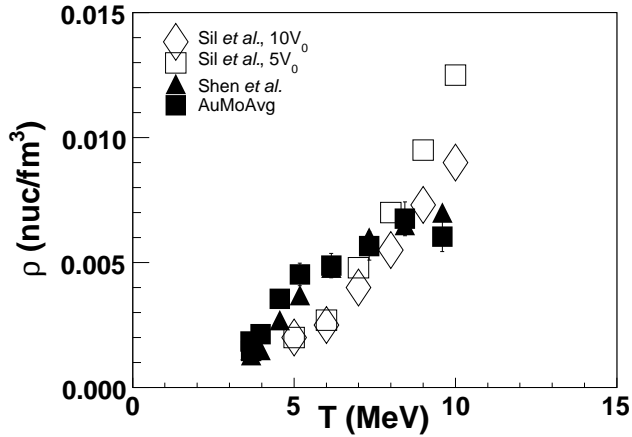,width=9.2cm,angle=0}
\caption{Estimated nuclear densities as a function of temperature. 
Values derived from the experimental yield ratios using the Alberga formulas
are indicated by solid squares.  Theoretical results from 
reference~\onlinecite{albergo85} are shown by solid triangles.  Values
derived from the theoretical calculations of
reference~\onlinecite{sil04} are
shown for confining volumes of 5V$_0$ (open squares) and
10V$_0$ (open diamonds).}
\label{fig5}
\end{figure}

Once  the free nucleon densities are known, the densities  of  the
other  particles  may  be  calculated  from  the  experimentally
observed  yields.  This, again is done as a  function  of  surface
velocity.  The  results,  obtained by  summing  the  densities  of
particles  with  A  =  1 to 4, for the two reaction  systems,  are
presented in Figure~\ref{fig5}.  The values for  the two systems show  quite
good  agreement and indicate low densities.  It  is  worth  noting  that  our
measurements of both the temperature and the associated alpha mass
fraction,  X$_\alpha$, also provide a means of estimating the densities  by
comparison with the Schwenk and Horowitz~\cite{horowitz05} or 
Shen {\it et al.}~\cite{shen98} calculations.  Like  the  Albergo calculation 
these  calculations assume chemical equilibrium and lead to similar low 
densities.

For the higher temperatures and densities, where the Shen {\it et al.}
calculations include competition from heavier species, the Shen {\it et al.}
calculation indicates slightly higher densities for a given temperature and 
alpha mass fraction.  In the following analysis we employ the densities from 
the Albergo model treatment. The resultant densities, presented in column 4 of
Table~\ref{tab1} and in Figure~\ref{fig5}, range from $1.5-7.0\times10^{-3}$
nucleons/fm$^{3}$.  For comparison to these values we also show in 
Figure~\ref{fig5}, theoretical results reported by
Sil {\it et al.}~\cite{sil04} for hot Thomas-Fermi model calculations for
$^{185}$Re confined in volumes of 5 or 10 times V$_{0}$, the nuclear volume
at normal density.  At temperatures of 5 to 10 MeV, they found average gas
(or low density surface matter) densities in the range of 2 to 
$10\times10^{-3}$ nucleons/fm$^{3}$.  Below T = 10 MeV these values are
in good agreement with those derived from comparison of the data
to the VEOS results and provide some support for those derived from the
experiment. In an earlier application of the virial equation
approach to similar types of reactions,  Pratt {\it et al.}~\cite{pratt87}
derived densities of $1-3\times 10^{-3}$ nucleons/fm$^3$ and temperatures
of 4.38 MeV to 7.90 MeV. These are in the range of the present
results.  Additionally, it is worth noting that the derived gas
densities at the temperatures in Table~\ref{tab1} are quite
similar to those indicated by recent Fisher scaling model analyses
of fragmentation data by Elliott {\it et al.}~\cite{elliott03}.

It is also possible to estimate gas densities using AMD-V model calculations 
once a calculational criterion capable of distinguishing gas nucleons from 
liquid nucleons is established.  For this calculation, we first verified
that the parameterization of the AMD-V model reproduced, at 300 fm/c, a
total mass fraction of gas, i.e. of particles with A$\leq4$, in reasonable
agreement with that derived from the experiment.  Although clusters are 
naturally produced in AMD-V calculations, early recognition of clusters 
relies on a physical space coalescence radius parameter.  The value of 2.5~fm 
was used in this calculation.  We then calculated, as a function of time,  the
average separation distance between all nucleons assigned to the gas.
From the average separation distance the average density is readily derived
assuming spherical symmetry in the source frame.  The AMD-V derived
densities, averaged for Mo and Au targets, are in the range of 2 to
$4\times10^{-3}$ nucleons/fm$^{3}$.  The agreement between the AMD-V results 
and these experimental values in Figure~\ref{fig5}, while not perfect, is 
reasonable, particularly when it is recognized that the AMD-V results demand 
a separation of nucleons for identification as ``gas'' while the clustering 
may actually occur in a slightly higher density surface region of matter 
identified as ``liquid'' by the AMD-V analysis.   Thus the results depend on 
the early recognition coalescence parameter.

\begin{table}[b]
\caption{}
\begin{center}
\begin{tabular}{|lllll|}
\hline
V$_{surf}$ & T     & F$_{sym}$ & $\rho$      & E$_{sym}$ \\
\hline
cm/ns      & MeV   & MeV       &  nuc/fm$^3$ &  MeV      \\
\hline
0.75       & 3.65 &  5.40     & 0.00184      &  9.03     \\
1.25       & 3.68 &  5.67     & 0.00147      &  9.66     \\
1.75       & 3.97 &  6.69     & 0.00213      &  11.1     \\
2.25       & 4.56 &  7.63     & 0.00355      &  11.9     \\
2.75       & 5.18 &  8.11     & 0.00453      &  12.1     \\
3.25       & 6.16 &  9.73     & 0.00488      &  13.6     \\
3.75       & 7.33 & 11.3      & 0.00566      &  13.6     \\
4.25       & 8.44 & 11.7      & 0.00675      &  13.4     \\
4.75       & 9.60 & 12.2      & 0.00604      &  12.8     \\
\hline
\end{tabular}
\end{center}
\label{tab1}
\end{table}

\section{Entropies and Symmetry Energies}
For the densities and temperatures derived from our data, we have
employed the entropy equations of the VEOS calculation to determine the
temperature and density dependent entropic contribution to F$_{sym}$.  
We have done this for the densities and temperatures of Table~\ref{tab1}.

Given  that  the  isoscaling parameter is sensitive  to  both  the
symmetry energy and symmetry entropy contributions, derivation  of
the  temperature and density dependent symmetry energy coefficient
from  the  isoscaling  parameter requires  an  evaluation  of  the
symmetry  entropy  contribution. We have based our  extraction  of
this  contribution  on the entropy equations of the Schwenk and Horowitz
paper. In that work, the entropy density is expressed as

\begin{eqnarray}
s &=& \frac{5P}{2T}-n_n\log{z_n} - n_p\log{z_p} - n_\alpha\log{z_\alpha} \\
&+&\frac{2T}{\lambda^3}\bigl((z_n^2+z_p^2)b_n' + 2z_pz_nb_{pn}'\bigr) \\
&+&\frac{T}{\lambda_\alpha^3}\bigl(z_\alpha^2b_\alpha' + 2z_\alpha(z_n+z_p)b_{\alpha n}'\bigr)
\end{eqnarray}

Where P is the pressure,  $n_n$, $n_p$ and $n_\alpha$ are the $n$, $p$ and 
$\alpha$-particle densities,  $z_n$, $z_p$  and alpha $z_\alpha$ are the 
particle fugacities, $\lambda$($\lambda_\alpha$) denotes the nucleon 
($\alpha$ particle) thermal wavelength  and the $b_i$ terms denote the 
temperature derivatives of the second virial coefficients~\cite{horowitz05}

The entropy per nucleon is then
\begin{equation}
\frac{S}{A} = \frac{s}{n_b}
\end{equation}
where $n_b$ is the total baryon density in the gas.  Given that $^3$H and
$^3$He are observed experimentally, we add their mixing entropy terms,
contributions of the type, $-n_i\ln{z_i}$, to the entropy density
equation.  At the low apparent density of the nuclear gas with
which we are dealing, we assume that the particle interaction
terms are very small and that differences between them for the two
different systems studied may be neglected.  Thus we truncate the
equation by dropping the interaction terms.  We then replace the
fugacities in the mixing entropy terms by fractional yields and
calculate the entropy per nucleon for the two systems.  With some
scatter they exhibit an essentially linear dependence on surface
velocity.  Thus, we have proceeded by making a linear fit to the
entropy per nucleon as a function of surface velocity for each
system.  We then extract the entropy difference, $\Delta S$, between the
two systems.

Knowing the total neutron and proton composition for the gas in
the two systems we extract the symmetry entropy coefficient,
$S_{sym}$,

\begin{equation}
S_{sym} =  \Delta S / ((N-Z)/A)^2
\end{equation}

\begin{figure}
\epsfig{file=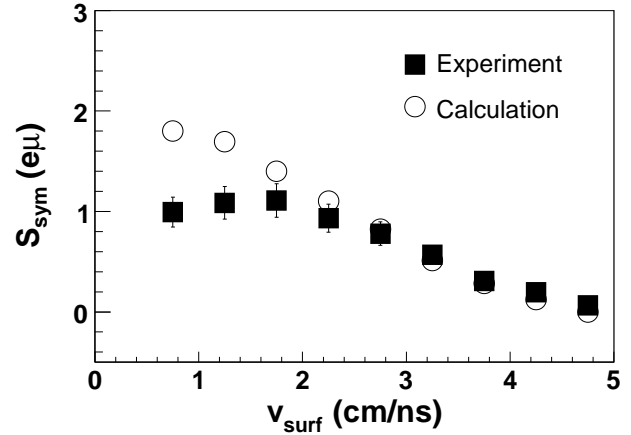,width=9.2cm,angle=0}
\caption{$S_{sym}$ vs $v_{surf}$.  Experimental data is represented by closed 
squares and calculated values are shown by open circles.}
\label{fig6}
\end{figure}

The values thus derived are presented as a function of surface
velocity in Figure~\ref{fig6}. For comparison we also show the values
calculated from the Schwenk and Horowitz VEOS for the
corresponding temperatures and densities. The agreement is quite
reasonable except at the very lowest density.

\begin{figure}
\epsfig{file=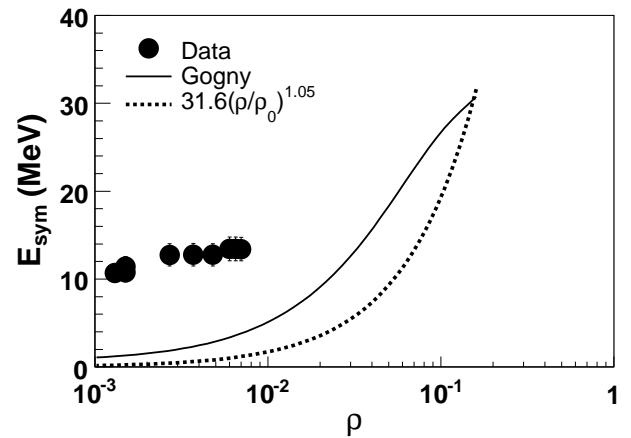,width=9.2cm,angle=0}
\caption{Derived symmetry energy coefficients as a function of
baryon density.  Solid diamonds indicate results using densities of
column 4 in Table~\ref{tab1}.  Solid line indicates the variation predicted
by the Gogny interaction.  The dotted line represents the function
$31.6(\rho/\rho_0)^{1.05}$~\cite{chen05}.}
\label{fig7}
\end{figure}

The symmetry energies obtained are presented in column 5 of Table~\ref{tab1}. 
The determination leads to values of the symmetry energy coefficients which 
range from 9.03 to 13.6 MeV.  We estimate the overall errors of these to be 
$\pm15\%$.  The symmetry energy coefficients are also plotted against density 
in Figure~\ref{fig7} where they are compared to those of uniform nuclear 
matter which are predicted by the Gogny effective 
interaction~\cite{decharge80} and to the $31.6(\rho/\rho_0)^{1.05}$ dependence
suggested by a recent analysis of isospin diffusion data~\cite{chen05}.  The
derived values of E$_{sym}$ are much higher than those predicted
by mean field calculations which ignore the cluster formation.

\section{Conclusions}
For nuclear gases with average proton fraction, $Y_p\sim0.44$, corresponding 
to the experimental average asymmetry of the Mo and Au target
results and densities at and below 0.05 times normal nuclear density, 
experimental symmetry energy coefficients of 9.03 to 13.6 MeV, with an 
estimated uncertainty of 15\% have been derived from experimentally
determined symmetry free energies, F$_{sym}$, determined using the
isoscaling method.  The symmetry energies are far above those
obtained in common effective interaction calculations and reflect cluster
formation, primarily of alpha particles, not included in such calculations.
The entropic contributions to the symmetry free energies derived from
isoscaling analyses are very important at low densities and are strongly 
affected by the cluster formation.The experimental values obtained agree 
well with the predictions of the Virial Equation of State model proposed by 
Horowitz and Schwenk~\cite{horowitz05}.  Inclusion of $^3$H, $^3$He and 
heavier clusters into the theoretical formalism is certainly desirable.  
Measurements at slightly higher densities should be more sensitive to the 
interaction terms in the VEOS.

\section{Acknowledgements}
This work was supported by the United States Department of Energy
under
Grant \# DE-FG03- 93ER40773 and by The Robert A. Welch Foundation
under
Grant \# A0330.  We appreciate very useful discussions with C.
Horowitz,
A. Schwenk and L. G. Moretto.

\end{document}